\documentclass[conference]{IEEEtran}
\IEEEoverridecommandlockouts
\usepackage[nospace]{cite}
\usepackage{amsmath,amssymb,amsfonts}
\allowdisplaybreaks[4]

\usepackage{algorithmic}
\usepackage{graphicx}
\usepackage{textcomp}
\usepackage{arydshln} % Add to preamble
\usepackage{xcolor}   % Add to preamble if not already there
\usepackage{booktabs} 
\usepackage{comment}

\def\BibTeX{{\rm B\kern-.05em{\sc i\kern-.025em b}\kern-.08em
    T\kern-.1667em\lower.7ex\hbox{E}\kern-.125emX}}

\makeatletter
\def\ps@IEEEtitlepagestyle{%
  \def\@oddhead{}%
  \def\@evenhead{}%
  \def\@oddfoot{%
    \textcolor{gray}{\scriptsize 979-8-3195-3554-2/26/\$31.00 ~\copyright2026 IEEE}%
    \hfil
  }%
  \def\@evenfoot{\@oddfoot}%
}
\makeatother

\begin{document}
\bstctlcite{IEEEexample:BSTcontrol}

\title{Comparing Contract-Based Support Mechanisms for Long-Duration Energy Storage
\thanks{
This work was funded by the Taylor Donation from the Grantham Institute and Energy Futures Lab, and Global Fellows Fund from Imperial College London,  the Leverhulme International Professorship with grant reference LIP-2020-002, and the Engineering and Physical Sciences Research Council under the grant EP/Y025946/1 (Electric Power Innovation for a Carbon-free Society (EPICS)).\\
Corresponding author: Adam Suski (e-mail:
a.suski23@imperial.ac.uk)
}
}
\author{
\IEEEauthorblockN{
Adam Suski\IEEEauthorrefmark{1},
Elina Spyrou\IEEEauthorrefmark{1},
Jacob Mays\IEEEauthorrefmark{2},
Richard Green\IEEEauthorrefmark{3}
}

\IEEEauthorblockA{\IEEEauthorrefmark{1}
Department of Electrical and Electronic Engineering, Imperial College London, London, United Kingdom
}

\IEEEauthorblockA{\IEEEauthorrefmark{2}
School of Civil and Environmental Engineering, Cornell University, Ithaca, NY, USA
}

\IEEEauthorblockA{\IEEEauthorrefmark{3}
Department of Economics \& Public Policy, Imperial Business School, London, United Kingdom
}
}

\maketitle

\begin{abstract}
Long-duration energy storage (LDES) faces significant revenue volatility that impedes investment. This paper evaluates four contract-based support mechanisms using an equilibrium model with risk-averse investors and incomplete risk markets. Applied to a stylized 2035 Great Britain case, we find that all mechanisms can achieve the targeted LDES capacity but differ substantially in cost-effectiveness and risk-aversion sensitivity. Contracts that eliminate revenue volatility achieve the lowest costs but may weaken operational incentives, while contracts that preserve market exposure maintain incentives at higher costs.
\end{abstract}


\section{Introduction}
Long-duration energy storage (LDES) is widely regarded as a critical technology for deep decarbonization of power systems \cite{sepulveda_design_2021, dowling_role_2020}. Long-term power system development plans worldwide, such as in Great Britain \cite{eso_future_2024} and Australia \cite{australia_energy_market_operator_aemo_2024_2024}, include a rapid expansion of LDES. Nevertheless, viable business models that allow investors to achieve their target returns with sufficient confidence have not yet been fully developed \cite{cheng_electricity_2025}.

A key challenge for LDES is the inherent volatility and uncertainty of its revenues \cite{suski_missing_2025}. This volatility primarily reflects dependence on (i) medium- to long-term weather patterns when variable renewable energy (VRE) dominates supply for charging and (ii) scarcity events in which LDES becomes the marginal producer, assuming the adequacy role traditionally fulfilled by gas-fired peakers and recovering its capital costs through high spot prices.

Given the substantial system benefits that LDES can deliver \cite{staadecker_value_2024, dowling_role_2020}, policymakers worldwide have proposed dedicated support mechanisms to incentivize deployment. Most designs aim to encourage competition \textit{for} long-term LDES contractual agreements and let assets compete \textit{in} the short-term energy market, leaning toward the concept of hybrid energy markets  \cite{roques_adapting_2017, joskow_hierarchies_2022}. The UK has launched a Cap-and-Floor (C\&F) scheme \cite{department_for_energy_security__net_zero_long_2024, spyrou_designing_2026}, modeled after its electricity interconnector framework \cite{ofgem_cap_2016}. Australia has initiated competitive tenders under the Long-Term Energy Service Agreement (LTESA) model \cite{nsw_department_of_planning_industry_and_environment_long-term_2021}.\footnote{The UK’s Cap and Floor scheme guarantees LDES a minimum net revenue floor for capital cost recovery at a notional rate while capping excess profits to protect consumers. Australia’s LTESA uses a similar options-based mechanism, providing a periodic floor while preserving market upside.} Similar programs are emerging in New York \cite{new_york_state_energy_research_and_development_authority_bulk_2026}.

However, the analytical support for designing these contracts has been confined to static financial assessments, such as that done by CEPA \cite{cepa_cap_2025} for the UK's scheme, which treat investment quantities as exogenous and are therefore structurally incapable of capturing the endogenous feedback between contract parametrization, investors' risk-adjusted required returns, and equilibrium capacity outcomes. In general, there is a lack of evidence evaluating whether a specific parametrization achieves the target capacity, generates fiscal overexposure, or induces market distortions. To assess the effectiveness of these mechanisms and their alternative parameterizations, models that capture the interactions among investment decisions and the mechanisms' impacts on investors' risk-adjusted profits and consumer surplus are necessary. The results of these models can provide policymakers and regulators with insights into the strengths and weaknesses of different options, while accounting for the second-order effects of mechanisms that drive the system to new equilibrium points.

In this context, this paper aims to compare alternative contract-based support mechanisms that encourage cost-effective LDES investment and manage LDES revenue risk, and to assess the sensitivity of contract performance to investors' risk preferences.  To achieve these objectives, we construct an equilibrium capacity-investment model featuring risk-averse LDES investors operating in incomplete risk markets with limited access to long-term hedging contracts \cite{tirole_incomplete_1999}. Within this modeling framework, we represent a series of policy-relevant contract designs, spanning different parameterizations and levels of risk aversion. We then compare their performance using system-level outcomes, equilibrium investment levels, and welfare metrics.
While alternative LDES support mechanisms have been widely discussed \cite{mastropietro_taxonomy_2024, twitchell_targeted_2025}, to our knowledge, this study offers the first equilibrium-based comparison of these contract designs in incomplete-risk markets.

\section{Methodology}

We build on the two-stage stochastic equilibrium model introduced by Mays et al. \cite{mays_asymmetric_2019} and extended in subsequent literature \cite{mays_financial_2023, shu_beyond_2023, risanger_congestion_2024}. This model is an instance of an incomplete capacity investment equilibrium problem, which is non-convex due to a non-monotone complementarity mapping arising from incomplete risk trading across agents \cite{ehrenmann_generation_2011, ralph_risk_2015, abada_multiplicity_2017}, and lacks an equivalent centralized optimization formulation. The model captures investment behavior under uncertainty for a set of resources $r \in \mathcal{R}$, including generators and energy storage, subject to predefined contract payoffs. Uncertainty is represented by a finite set of scenarios $\omega \in \Omega$, each assigned a probability $p_\omega$. In the following subsections, we present the abstract formulation for both the second and first stages. The detailed formulations of the objective functions and feasibility sets for the second-stage model follow largely those in \cite{suski_missing_2025} and \cite{mays_asymmetric_2019}.

\subsection{Second-stage}
The second stage occurs after the uncertainty has cleared and investment has been fixed. Hence, the problem $\text{DIS}_\omega$ maximizes the scenario-specific welfare $U_\omega$ by deciding on demand $d_{\omega t}$ and the dispatch of resources $q_{\omega t r}$ in each time period $t \in \mathcal{T}$ in each scenario $\omega \in \Omega$:
\begin{subequations}
\label{eq:Dispatch}
\begin{gather}
\text{DIS}_\omega:  \max  U_\omega = \sum_{t \in \mathcal{T}} \left(U_{\omega t}^c(d_{\omega t}) - \sum_{r \in \mathcal{R}}  C_{\omega r}(q_{\omega t r}) \right)\\
\sum_{r \in \mathcal{R}} q_{\omega t r} = d_{\omega t}, \; (\lambda_{\omega t}) \quad \forall t \in \mathcal{T} \\
d_{\omega t} \in \mathcal{Y}^d \quad \forall t \in \mathcal{T}\  \\
q_{\omega t r} \in \mathcal{Y}_r \quad \forall r \in \mathcal{R},\ t \in \mathcal{T}\ \\
q_{\omega t r}\geq 0,  d_{\omega t}\geq 0 \quad \forall r \in \mathcal{R},\ t \in \mathcal{T} 
\end{gather}
\end{subequations}

Here, $d_{\omega t}$ denotes demand, $q_{\omega t r}$ the dispatch of resource $r$, $U_{\omega t}^c(\cdot)$ the consumer hourly utility function, and $C_{\omega r}(\cdot)$ the total hourly operating cost function of resource $r$. Sets $\mathcal{Y}^d$ and $\mathcal{Y}_r$ define the feasible operational regions for demand and each resource, respectively. Dual variables $\lambda_{\omega t}$, corresponding to the balancing constraint, represent electricity market prices and are used to compute scenario-specific risk-adjusted profits and price-based contract payoffs.

\subsection{First-stage}
In the first stage, each investor solves problem $\text{INV}_r$, maximizing its risk-adjusted profit $\rho_r$, which is represented as a convex combination of the expected value and Conditional Value at Risk ($\text{CVaR}_r$) of scenario net cashflows $u_{\omega r}$, where $\delta_r$ denotes the weight of the expected value and $\psi_r$ denotes the CVaR tail probability, following classical representation of \cite{rockafellar_optimization_2000}. Decision variables at this stage are investment levels $c_r$.
\begin{subequations}
\label{eq:Actor}
\begin{gather}
\text{INV}_r:\;
\max_{}
\rho_r = (1-\delta_r) \text{CVaR}_r + \delta_r \sum_{\omega\in\Omega} p_\omega u_{\omega r},\\
\text{CVaR}_r = 
\zeta_r
-
\frac{1}{\psi_r}
\sum_{\omega\in\Omega} p_\omega u^+_{\omega r}
\\
u_{\omega r} = \pi_{\omega r} c_r - C_r^{\mathrm{I}} c_r + c_r \kappa_{\omega r}, \\
u^+_{\omega r} \ge \zeta_r - u_{\omega r},
\qquad
u^+_{\omega r} \ge 0,
\quad \forall \omega\in\Omega
\end{gather}
\end{subequations}

Here, $\zeta_r$ represents Value at Risk (VaR), $u^+_{\omega r}$ is the CVaR auxiliary variable, and $u_{\omega r}$ is the scenario-specific net cashflow of resource $r$. The term $\pi_{\omega r}$ denotes the scenario-specific market net revenue per unit of installed capacity, computed from second-stage dispatch outcomes and market prices, net of operating costs. $C_r^{\mathrm{I}}$ denotes the investment cost per unit of capacity, while $\kappa_{\omega r}$ denotes the contract payoff per unit of capacity, which can be a cash inflow ($\kappa_{\omega r}>0$) or outflow ($\kappa_{\omega r}<0$) depending on the scenario and contract structure.

Note that, unlike in \cite{mays_asymmetric_2019}, we assume that specified resources can be built only under contracts and that such contracts cover their total installed capacity. This represents a case in which contracts are centrally allocated, often by state agencies, with predefined characteristics and a zero contract premium, implying that the mechanism aims to transfer revenue risk without charging a premium to the asset in the spirit of Laffont and Tirole \cite{laffont_using_1986}. This approach aligns with current practice in GB, where the regulator allocates contracts based on an assessment of multi-attribute bids \cite{ofgem_long_2025}.

\subsection{Solution strategy}

Following \cite{shu_beyond_2023}, the long-run equilibrium constraints for each investor in \eqref{EQ:InvestmentCondition} require zero risk-adjusted profit in a market with free entry and exit, defined through a bounded complementarity condition \cite{steven_a_gabriel_complementarity_2013}:
\begin{gather} 
0 \le c_r \le \bar c_r \;\perp\; \rho_r \geq 0, \quad \forall r\in\mathcal{R}. \label{EQ:InvestmentCondition} 
\end{gather}

where $\bar c_r$ denotes the maximum investable capacity of resource $r$. 

Unlike the risk-neutral or complete-market setup \cite{ralph_risk_2015}, no straightforward centralized-optimization formulation of this problem exists. Alternative methods employ reformulations \cite{dimanchev_consequences_2024, byers_electricity_2025} or algorithms such as ADMM \cite{hoschle_admm-based_2018} or variants of the Gauss-Seidel method \cite{mays_asymmetric_2019}. In this work, our set of assumptions allows us to use a simple procedure to find equilibria. In all experiments, only a single contract type is available to a single technology, hence, contract payoffs are indexed only by scenario and resource. We consider multiple discrete levels of technology capacity that qualify for the contract. For each capacity level, we find the capacity levels for the remaining technologies that yield zero risk-adjusted profit. For the qualifying technology, the equilibrium capacity is set to yield zero risk-adjusted profits when contract payoffs are included. In summary, the search identifies fixed points of the complementarity condition \eqref{EQ:InvestmentCondition} over the discretized capacity domain of the technology qualifying for the contract.
\section{Contract definitions} \label{SEC: ContractDefinition}

In this section, we define the contract types considered in the analysis and their corresponding payoff functions.

%Scenario-based payouts are represented by $\kappa_{\omega r}$, where a positive value indicates a cash flow to the asset and a negative value indicates a cash flow from the asset. 
\subsection{Cap and floor}

A Cap and Floor (C\&F) contract, also called a collar \cite{billimoria_contract_2023}, guarantees a minimum level of periodic net revenue to ensure bankability and fundability while protecting consumers from high costs by limiting excessive profits. Let $\hat{\pi}^F_r$ and $\hat{\pi}^C_r$ denote the floor and cap revenue levels, respectively, with $\hat{\pi}^C_r \ge \hat{\pi}^F_r$. The scenario-based payoff is given by
\begin{gather}
\kappa_{\omega r}
= \max\!\left(\hat{\pi}^F_r - \pi_{\omega r},\,0\right) - \max\!\left(\pi_{\omega r} - \hat{\pi}^C_r,\,0\right).
\end{gather}

Here, $\hat{\pi}^F_r$ and $\hat{\pi}^C_r$ are determined by applying the specified cost-of-capital rates (i.e., the rates used in the capital recovery factor over the asset lifetime) to the fixed costs, thereby establishing the absolute floor and cap revenue levels.

\subsection{Revenue CfD}
The net Revenue Contract for Difference (R-CfD), also called a revenue swap \cite{billimoria_contract_2023}, stabilizes asset income by guaranteeing a revenue level over a multi-year contract period. The asset owner receives a top-up payment when net revenue falls below the strike revenue and pays back excess revenue when it exceeds the strike level. It is a special case of C\&F in which the cap and the floor are identical. The payoff is calculated as
\begin{equation}
    \kappa_{\omega r} = \hat{\pi}_r - \pi_{\omega r}.
\end{equation}

\subsection{Spread CfD}
The Spread CfD (S-CfD) stabilizes storage asset revenues by guaranteeing a price spread between charging and discharging, scaled by the total discharged volume, over a multi-year contract. This type of contract was recently proposed in \cite{mastropietro_taxonomy_2024}. The asset owner receives a payment when its scenario-specific realized spread $\sigma_{\omega r}$ falls below the strike spread $\hat{\sigma}_r$ and pays back excess revenue when it exceeds the strike level:
\begin{equation}
    \kappa_{\omega r} = (\hat{\sigma}_r - \sigma_{\omega r}) v_{\omega r}
\end{equation}

Here, $\sigma_{\omega r}$ is the realized spread and $v_{\omega r}$ is the electricity discharged normalized by installed capacity.

\subsection{Availability Contract}

The Availability Contract (AvC) is a fixed cash inflow for the storage asset, scaled by the unit's availability. The asset owner receives a fixed payment rate ($\hat{a}_r$) for available capacity. 
\begin{equation}
\kappa_{\omega r} = \tau_{\omega r} \hat{a}_r,
\end{equation}
$\tau_{\omega r}$ is the share of hours in the year with the storage asset available. Following \cite{billimoria_contract_2023}, we consider a unit available during any hour $t$ in which the storage asset is either charging or discharging, and unavailable otherwise. 

\section{Case study}

The case study represents a stylized Great Britain power system in 2035, with fixed capacities of 35~GW unabated combined-cycle gas turbines (CCGT) and 5~GW nuclear. Investment in renewables and storage is decided by the model. Storage options include 2-hour batteries and 12-hour LDES; renewable technologies comprise solar PV, onshore wind (capped at 30~GW), and offshore wind. Peak demand is 65~GW \cite{eso_future_2024}, the price cap is 20{,}000~\$/MWh, and flexible demand (following the formulation in \cite{suski_missing_2025}) is fixed at 2~GW. Uncertainty is captured by 60 equiprobable scenarios combining 20 weather realizations with three gas cost levels (78, 150, 220~\$/MWh). The model spans one year with 1{,}440 representative time steps per scenario \cite{pineda_chronological_2018}. LDES investment is undertaken by a risk-averse investor (CVaR with $\psi=0.2$, varying $\delta_r$ for sensitivity). Other investors are risk-neutral.\footnote{We assume that existing financial markets and government support mechanisms adequately de-risk investments in more established technologies, which may also access revenue streams from ancillary services not explicitly modeled here.} Risk-free Weighted Average Cost of Capital (WACC) for LDES is 7.1\%.

The foundation for our mechanism assessment is the set of \emph{net-zero and regulatory trilemmas} outlined by Pollitt et al. \cite{pollitt_uncertainty_2025}, specifically effectiveness (whether the regulation achieves its intended outcome), responsiveness (whether it distorts other aspects of behaviour), and coherence (whether it will be seen as consistent and predictable), concentrating on the first of these.
\section{Results}

\subsection{Impact of the lack of long-term hedging contracts}

%While several studies have estimated the impact of the lack of hedging instruments (e.g. \cite{makrides_quantifying_2025} and \cite{dimanchev_consequences_2024}), we extend the literature by directly linking these effects to changes in financing costs.

Table~\ref{tab:wacc_delta_selected} reports changes in selected parameters across different levels of risk aversion. We use implied WACC (or the associated risk premium) as a summary metric for financing conditions driven by investor risk aversion, as it provides an interpretable lens on the system-wide impacts of incomplete risk markets \cite{ehrenmann_generation_2011}.\footnote{We estimate the implied WACC (a proxy for financing cost) using \cite{mays_financial_2023}, as the discount rate at which the present value of mean annual net revenue (net of fixed O\&M and variable costs) equals the upfront investment cost annualized at the risk-free rate.} LDES capacity declines as LDES risk aversion increases, with $\delta = 0.6$ leading to a reduction of more than 50\% in capacity and an almost 3 percentage point increase in financing costs. It is associated with an endogenous increase in the required return due to higher sensitivity to downside risk. Consistent with \cite{mays_financial_2023} and \cite{dimanchev_consequences_2024}, both prices and adequacy are adversely affected. CCGT generation decreases as the VRE and storage mix change, as financing costs for LDES rise.

\begin{table}[h]
\centering
\scriptsize
\setlength{\tabcolsep}{4pt}
\caption{Selected equilibrium outcomes across risk-aversion levels}
\label{tab:wacc_delta_selected}
\begin{tabular}{lccccc}
\toprule
\textbf{Metric} & \multicolumn{5}{c}{\textbf{Risk aversion parameter } $\boldsymbol{\delta}$} \\
\cmidrule(lr){2-6}
 & 1.0 & 0.9 & 0.8 & 0.7 & 0.6 \\
\midrule
Implied WACC (\%)                 & 7.1  & 7.8   & 8.4   & 9.1   & 9.8   \\
LDES capacity (GW)                & 16.02 & 14.64 & 13.30 & 11.13 & 7.26  \\
Avg. electricity price (\$/MWh)   & 269.9 & 276.0 & 282.1 & 287.7 & 293.1 \\
Exp. unmet demand (GWh)          & 101.6 & 106.0 & 112.1 & 117.9 & 125.0 \\
Exp. generation CCGT (TWh)        & 58.7  & 58.1  & 57.7  & 57.1  & 56.2  \\
Exp. generation VRE (TWh)         & 294.1 & 294.3 & 294.3 & 294.1 & 293.8 \\
Exp. VRE curtailment (TWh)        & 45.0  & 47.1  & 49.0  & 50.9  & 53.2  \\
\bottomrule
\end{tabular}
\end{table}

\subsection{Contract impact on LDES revenue distribution} \label{SEC: RevDist}

\begin{figure}[!htbp]
    \centering
    \includegraphics[width=\linewidth]{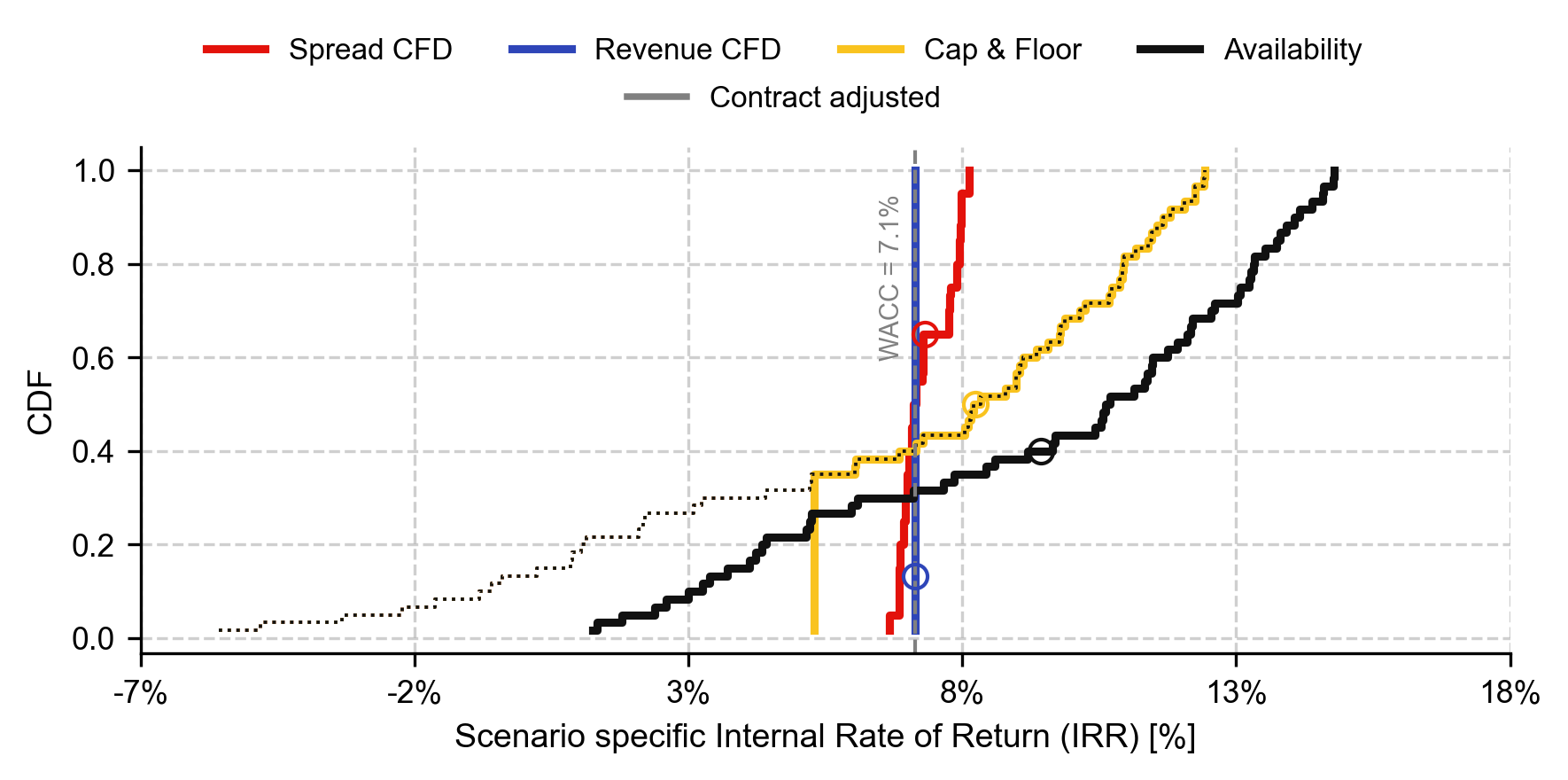}
    \caption{Empirical CDFs of scenario‑specific IRR for contract parametrizations yielding risk-neutral capacity (16~GW). Solid and dotted lines show the IRR with and without contract payoffs, respectively. Round markers denote the expected value of the IRR (interpreted as implied WACC \cite{mays_financial_2023}).
}
    \label{fig:irr_cdfs_contract_scenarios_single_parameterization}
\end{figure}

Here, we assess how alternative contract-based support mechanisms shape the distribution of LDES revenues. Figure~\ref{fig:irr_cdfs_contract_scenarios_single_parameterization} presents the cumulative distribution functions of scenario-specific IRRs under all mechanisms. The R-CfD collapses to a single vertical line, fully eliminating net revenue volatility. Owing to this lack of distributional exposure, it has the same impact regardless of the investor’s risk aversion, thereby incentivizing a target level of LDES capacity. Under the S-CfD, revenue volatility arises from differences across scenarios in the electricity discharged. For 12-hour storage, these differences are relatively small across scenarios, resulting in a still-narrow distribution of net revenues after contract adjustment. The C\&F truncates revenues outside the collar;\footnote{The 14\% cap is not binding, so its impact is not visible in Figure \ref{fig:irr_cdfs_contract_scenarios_single_parameterization}.} therefore, the relative positioning of the cap and floor determines LDES investment. As with electricity discharge, availability (as defined above) exhibits low variability across scenarios, yielding nearly constant AvC payoffs. Consequently, the contract-adjusted revenue distribution under the AvC retains the shape of the uncontracted distribution, shifted rightward by an amount that varies across scenarios due to the nonlinearity of IRR with respect to cash flows, with the shift being proportionally larger at lower revenue levels.

Table~\ref{tab:risk_adjusted_revenues} summarises revenue statistics for LDES under all mechanisms. By collapsing the revenue distribution, R-CfD and S-CfD eliminate variability, as reflected in the implied WACC, which is estimated at or close to the risk-free rate of 7.1\%. AvC, while restoring risk-neutral capacity, does not materially narrow the revenue distribution; consequently, it yields only a 0.3 percentage point reduction in implied WACC relative to the fully incomplete case (Table~\ref{tab:wacc_delta_selected}). In contrast, C\&F improves the implied WACC by truncating the lower tail of the revenue distribution, reducing it by 1.6 percentage points relative to the fully incomplete case. However, it remains approximately 1 percentage point above the risk-free benchmark.

\begin{table}[!htb]
\caption{Annual revenues statistics and implied financing costs at the target LDES capacity ($\delta_r = 0.6$).}
\label{tab:risk_adjusted_revenues}
\centering
\scriptsize
\renewcommand{\arraystretch}{0.95}
\setlength{\tabcolsep}{2.5pt}
\begin{tabular}{lccccc}
\toprule
Mechanism & CV [-]& MIN [\%] & CVaR$_{0.2}$ [\%] & Implied WACC [\%]\\
\midrule
Risk neutral & 49.2 & -5.6 & -1.0 & - \\
AvC  & 37.9 & 1.3 & 3.0 & 9.5 \\
C\&F  & 27.2 & 5.3 & 5.3 & 8.2 \\
R-CfD  & 0.0 & 7.1 & 7.1 & 7.1 \\
S-CfD  & 5.4 & 6.7 & 6.8 & 7.3 \\
\bottomrule
\end{tabular}

\vspace{2pt}

\scriptsize{CV - coefficient of variation. MIN is the minimum revenue observed across scenarios, and CVaR$_{0.2}$ is the average revenue of the worst 20\% of scenarios. MIN and CVaR are reported as IRR.  Implied WACC is interpreted as the expected IRR from Figure \ref{fig:irr_cdfs_contract_scenarios_single_parameterization}.}
\end{table}

\subsection{Costs and consumer impacts of mechanisms studied} \label{SEC: InvestmentsUnderContracts}

With the right parametrization, all contracts can support the risk-neutral LDES capacity\footnote{Depending on the degree of market incompleteness and consumer risk aversion, a regulator may intend to target an alternative capacity level, for example, one corresponding to complete risk markets; see \cite{shu_beyond_2023}.} of 16~GW (Table~\ref{tab:wacc_delta_selected}). However, each mechanism entails different levels of risk and costs borne by consumers. Table~\ref{tab:mechanism_costs} reports mechanism costs, measured as expected contract payoffs per unit of total installed LDES, as well as per unit of additionally incentivized LDES relative to the incomplete-market outcome of 7.26~GW (Table~\ref{tab:wacc_delta_selected}).

\begin{table}[!htb]
\caption{Cost of alternative LDES support mechanisms at the target capacity ($\delta_r = 0.6$).}
\label{tab:mechanism_costs}
\centering
\scriptsize
\renewcommand{\arraystretch}{0.95}
\setlength{\tabcolsep}{4pt}
\begin{tabular}{l|c|c|c|c}
\toprule
 & AvC & C\&F & R-CfD & S-CfD \\
\midrule
Optimal mechanism parametrization & 31.1 & 5.3 & 100.0 & 463.7 \\
Mechanism cost per MW (installed) & 30 & 14 & 0 & 2 \\
Mechanism cost per MW (incentivized) & 54 & 25 & 0 & 4 \\
\bottomrule
\end{tabular}
\vspace{2pt}

\scriptsize{Parameterization follows all ($\hat{\cdot}$) symbols in Section~\ref{SEC: ContractDefinition}: AvC and R-CfD values are given as \% of annualized fixed cost, C\&F strikes as the floor level with a 14\% cap and S-CfD strikes in \$/MWh. Mechanism costs are shown as \% of annualized fixed cost per MW. ``Installed'' refers to total LDES capacity (16~GW), while ``incentivized'' refers to additional capacity beyond the incomplete-market level (7.26~GW).}\end{table}

The R-CfD bears zero cost, as expected, since it fully removes revenue variability and, under risk-neutral capacity, results in payments to and from the asset balance in expectation.\footnote{Zero expected net cost does not preclude administrative, settlement, and liquidity management costs borne by the counterparty.} In contrast, the costs of mechanisms that preserve some degree of revenue variability depend on the interaction between payoff structures and investors' risk profiles (and the extent of risk aversion, which we show in the next section). Accordingly, the S-CfD exhibits relatively low costs, followed by the C\&F. The highest costs are observed under the AvC, whose payouts are not directly linked to revenues and exhibit rather constant payouts across scenarios.

\begin{figure}[!htbp]
    \centering
    \includegraphics[width=\linewidth]{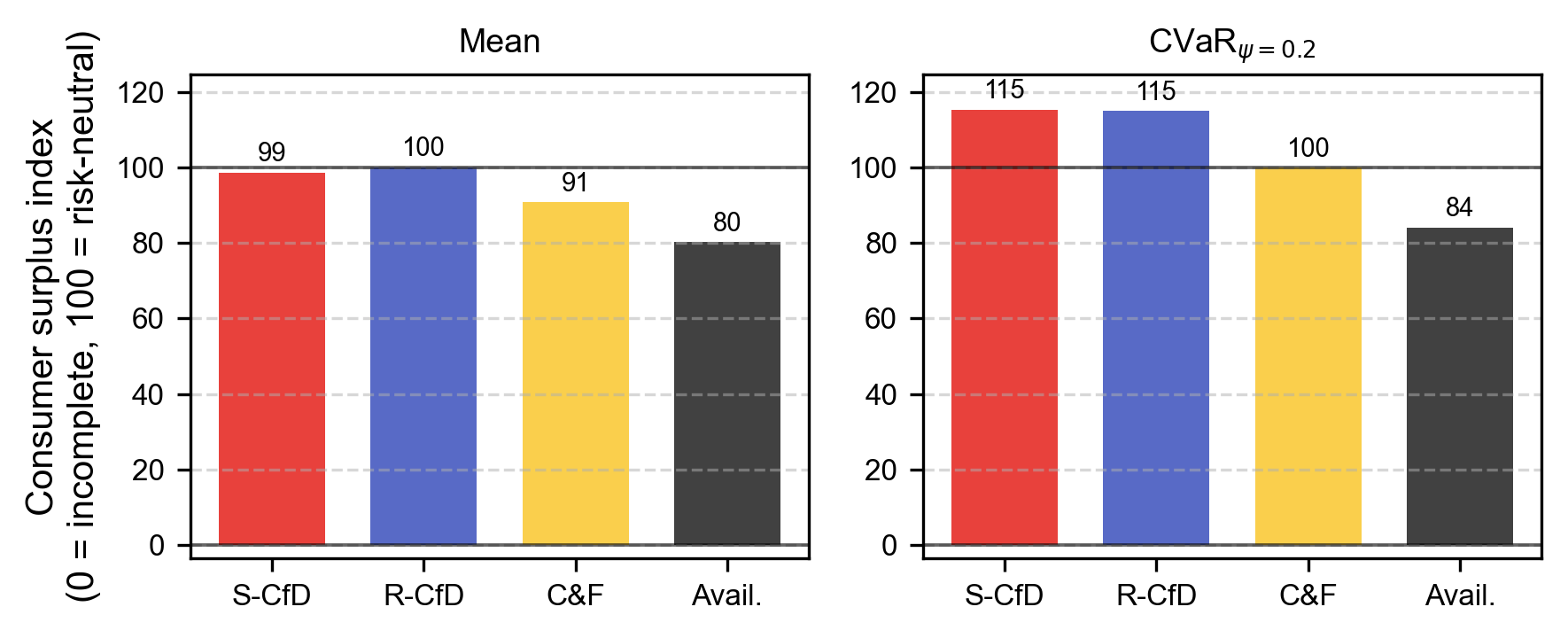}
    \caption{Relative consumer surplus performance of contract-based support mechanisms.  Each bar shows the mechanism-specific consumer surplus index 
($0 = \text{incomplete baseline}$, $100 = \text{risk-neutral baseline}$), under mean outcomes and downside-risk outcomes measured by $\mathrm{CVaR}$, with tail probability of ${\psi = 0.2}$.}
    \label{fig:consumer_welfare_mechanism_normalized_bars}
\end{figure}

In Figure \ref{fig:consumer_welfare_mechanism_normalized_bars} we show normalized levels of the mean and CVaR of consumer surplus. The R-CfD and S-CfD not only restore risk-neutral capacity but also increase consumer surplus when consumers are risk-averse. By contrast, under the C\&F, the cap is rarely binding, and the contract has limited interaction with consumer outcomes in tail scenarios; as a result, consumer CVaR remains close to the risk-neutral benchmark.

\begin{figure}[!htbp]
    \centering
    \includegraphics[width=\linewidth]{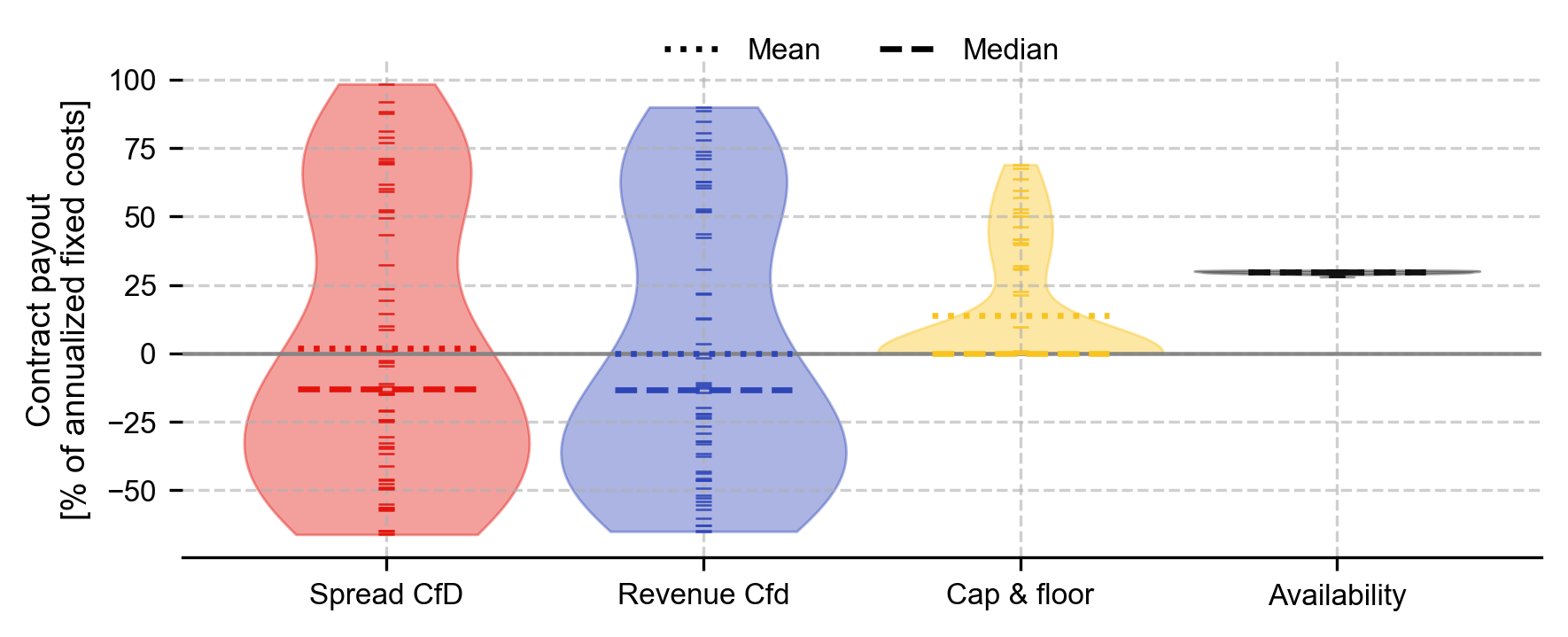}
    \caption{Distribution of scenario-level contract payouts as a share of annualized fixed costs. Short horizontal ticks indicate scenario realizations, with dotted and dashed lines denoting mean and median, respectively.}
    \label{fig:normalized_payment_scenario_violins}
\end{figure}

Figure~\ref{fig:normalized_payment_scenario_violins} shows contracts' payoff distributions normalized by annualized fixed cost. While R-CfD and S-CfD are cost-effective in expected net terms, with near-zero mean net payoffs reflecting the near balance of top-up and clawback flows, they exhibit the highest payoff volatility and the largest observable absolute payoffs. In contrast, AvC payoffs are unidirectional and, although their expected value exceeds that under the C\&F, are relatively predictable across scenarios, as reflected by their narrow distribution.

\subsection{Impact of varying risk aversion} \label{SEC: RiskAversionSenstivitiy}

In this section, we conduct a sensitivity analysis of contract performance with respect to investor risk aversion. We focus on two metrics: mechanism cost (Table~\ref{tab:mechanism_costs}) and implied WACC (Table~\ref{tab:risk_adjusted_revenues}). As expected, higher risk aversion (lower $\delta_r$) increases both mechanism costs and financing costs for all contracts except the R-CfD. For the S-CfD, costs rise primarily because the strike spread required to restore risk-neutral capacity increases with risk aversion, thereby modestly widening the revenue distribution. The AvC is the most sensitive to changes in risk aversion, exhibiting little resilience and a near-linear increase in costs. In contrast, the C\&F, which primarily insures the lower tail of the revenue distribution, shows a smaller increase in costs than the AvC as risk aversion rises, though a larger one than either CfD.

\begin{figure}[!htbp]
    \centering
    \includegraphics[width=\linewidth]{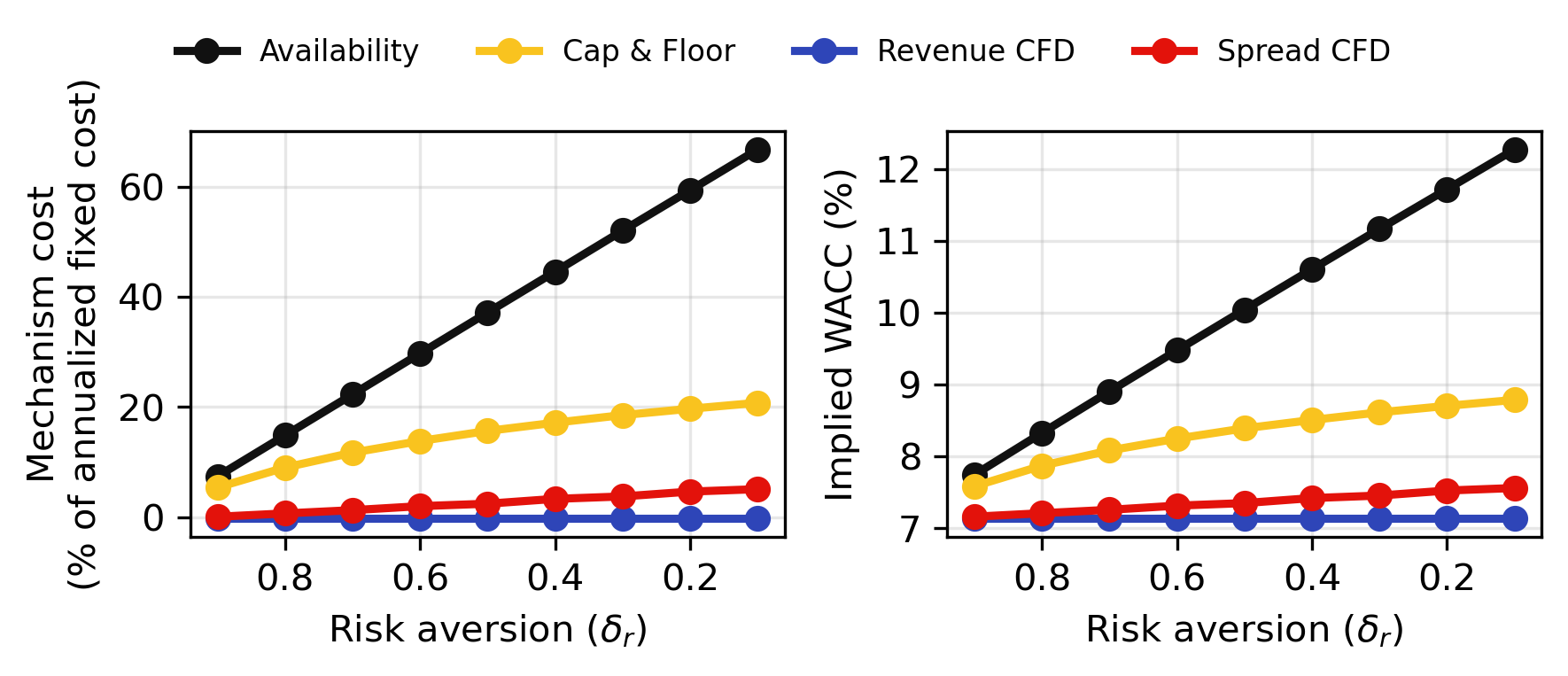}
    \caption{Mechanism cost (left) and implied LDES WACC (right) for varying risk aversion of the LDES investor under alternative contracts. All results are presented for contracts that recover the risk-neutral capacity of 16~GW.}
    \label{fig:mechanism_cost_and_wacc_vs_delta}
\end{figure}

\begin{comment}
\subsection{Impact of techno-economic parameters} \label{SEC: EfficiencySensitivity}

Varying techno-economic parameters can alter the relative performance of contract designs. For LDES, one widely discussed parameter is round-trip efficiency (RTE). Figure~\ref{fig:efficiency_threshold_capacity_payments_per_mw_double_subplot} varies RTE and examines the cost of each mechanism to restore risk-neutral capacity (which changes under different RTEs) and constant capacity of 16GW. We see that the overall mechanism cost is relatively insensitive to changes in RTE when the goal is to recover risk-neutral capacity. When capacity is set differently from the risk-neutral level, costs clearly increase, with primarly C\&F costs merging with R-CfD and S-CfD as the revenue distribution shifts to the left (due to excess capacity relative to the risk-neutral level).

\begin{figure}[!htbp]
    \centering
    \includegraphics[width=\linewidth]{figures/efficiency_threshold_capacity_payments_per_mw_double_subplot.png}
    \caption{Mechanism cost change with varying round-trip efficiency of LDES. The mechanisms are assessed to recover risk-neutral capacity, which varies across RTEs (left panel) and is constant at 16 GW (right panel). 64\% is the original level. 
}
    \label{fig:efficiency_threshold_capacity_payments_per_mw_double_subplot}
\end{figure}
\end{comment}

\section{Conclusions and discussion}

In this paper, we apply an equilibrium capacity-investment model with individual risk aversion to evaluate four contract-based LDES support mechanisms across a wide range of parametrizations and risk-aversion levels in a stylized 2035 Great Britain case, and identify equilibria via the zero risk-adjusted profit condition.

We find that for all mechanisms, there exists a set of parameters that achieves the target capacity. Our analysis abstracts from strategic dispatch distortions induced by contract design and focuses on long-run investment incentives. The net revenue contract-for-difference has the lowest expected cost (zero for risk-neutral capacity) because it eliminates net revenue volatility. The spread contract-for-difference also limits downside exposure and strongly compresses the revenue distribution, thereby reducing mechanism costs. Their performance is also largely resilient to changes in investor risk aversion because the residual net revenue volatility to which the investor remains exposed is zero or very low.

However, contracts that extensively insure revenue risk introduce several structural challenges. When parameters are administratively defined, such contracts may overcompensate some projects while excluding others, thereby shifting the regulator's role from price discovery to explicit capacity selection. Furthermore, as discussed extensively in the context of C\&F consultations \cite{ofgem_long_2025}, revenues that are decoupled from market outcomes weaken performance and dispatch incentives, introduce the risk of moral hazard, and require explicit mechanisms to ensure competitive and socially optimal operation \cite{fabra_designing_2025,billimoria_contract_2023}. Under the cap-and-floor and availability contracts, storage can remain more incentivized to maximize its market-related net revenue, thereby partially mitigating incentive-related concerns. However, allowing assets to be partially exposed to revenue risk reduces the cost-effectiveness of these mechanisms and yields smaller reductions in financing costs. Their design also involves more parameters. The availability contract results further indicate that availability, when measured by hours with non-zero output, may not be an effective indexing parameter for contract design, and that contract payoffs may reduce consumer surplus in tail scenarios relative to the risk-neutral target.

We also find the distribution of scenario-based contract payoffs substantially different across mechanisms. These differences can have policy-relevant implications: contracts with high gross-payoff volatility, even when associated with low expected net costs, require the offtaker to maintain liquidity buffers and establish appropriate credit arrangements, which may add to the overall cost of the mechanism and increase the administrative burden.

Taken together, these results and the proposed evaluation framework provide regulators with both tools and insights to inform the design of LDES support mechanisms as electricity markets continue to transform, integrating higher shares of variable renewable resources.

%\section*{Acknowledgments}

%\small This work was funded by the Taylor Donation from the Grantham Institute and Energy Futures Lab, Imperial College London, the Leverhulme International Professorship with grant reference LIP-2020-002, and the Engineering and Physical Sciences Research Council under the grant EP/Y025946/1 (Electric Power Innovation for a Carbon-free Society (EPICS)).

\newpage
% --- Bibliography (BibTeX) ---
\bibliographystyle{IEEEtran}
\bibliography{references2, references}

@misc{new_york_state_energy_research_and_development_authority_bulk_2026,
	title = {Bulk {Energy} {Storage} {Program}},
	url = {https://www.nyserda.ny.gov/All-Programs/Energy-Storage-Program/Developers-and-Contractors/Bulk-Storage-Incentives},
	language = {en},
	urldate = {2026-05-15},
	author = {{New York State Energy Research and Development Authority}},
	year = {2026},
}

@article{spyrou_designing_2026,
	title = {Designing a {Policy} {Mechanism} for {Long}-{Duration} {Energy} {Storage}: {The} {British} {Experience}},
	volume = {13},
	issn = {2196-3010},
	shorttitle = {Designing a {Policy} {Mechanism} for {Long}-{Duration} {Energy} {Storage}},
	url = {https://doi.org/10.1007/s40518-026-00287-y},
	doi = {10.1007/s40518-026-00287-y},
	language = {en},
	number = {1},
	urldate = {2026-04-08},
	journal = {Current Sustainable/Renewable Energy Reports},
	author = {Spyrou, Elina and Suski, Adam and Green, Richard},
	month = apr,
	year = {2026},
	pages = {5},
}

@article{suski_missing_2025,
	title = {Missing {Money} and {Market}-{Based} {Adequacy} in {Deeply} {Decarbonized} {Power} {Systems} with {Long}-{Duration} {Energy} {Storage}},
	issn = {2771-9626},
	url = {https://ieeexplore.ieee.org/document/11316365},
	doi = {10.1109/TEMPR.2025.3648914},
	urldate = {2026-02-23},
	journal = {IEEE Transactions on Energy Markets, Policy and Regulation},
	author = {Suski, Adam and Spyrou, Elina and Green, Richard},
	year = {2025},
	pages = {1--13},
}

@article{pineda_chronological_2018,
	title = {Chronological {Time}-{Period} {Clustering} for {Optimal} {Capacity} {Expansion} {Planning} {With} {Storage}},
	volume = {33},
	issn = {1558-0679},
	url = {https://ieeexplore.ieee.org/document/8369128/?arnumber=8369128},
	doi = {10.1109/TPWRS.2018.2842093},
	number = {6},
	urldate = {2024-10-08},
	journal = {IEEE Transactions on Power Systems},
	author = {Pineda, Salvador and Morales, Juan M.},
	month = nov,
	year = {2018},
	pages = {7162--7170},
}

@article{fabra_designing_2025,
	series = {51st {Annual} {Conference} of {European} {Association} for {Research} in {Industrial} {Economics}, {Amsterdam}, 2024},
	title = {Designing contracts for the energy transition},
	volume = {102},
	issn = {0167-7187},
	url = {https://www.sciencedirect.com/science/article/pii/S0167718725000396},
	doi = {10.1016/j.ijindorg.2025.103173},
	urldate = {2026-03-05},
	journal = {International Journal of Industrial Organization},
	author = {Fabra, Natalia and Llobet, Gerard},
	month = sep,
	year = {2025},
	pages = {103173},
}

@misc{byers_electricity_2025,
	address = {Rochester, NY},
	type = {{SSRN} {Scholarly} {Paper}},
	title = {Electricity {Market} {Design} and {Risk} {Trading} with {Flexible} and {Endogenous} {Demand}},
	url = {https://papers.ssrn.com/abstract=5953921},
	doi = {10.2139/ssrn.5953921},
	language = {en},
	urldate = {2026-02-25},
	publisher = {Social Science Research Network},
	author = {Byers, Conleigh and Billimoria, Farhad},
	month = dec,
	year = {2025},
}

@article{hoschle_admm-based_2018,
	title = {An {ADMM}-{Based} {Method} for {Computing} {Risk}-{Averse} {Equilibrium} in {Capacity} {Markets}},
	volume = {33},
	issn = {1558-0679},
	url = {https://ieeexplore.ieee.org/document/8294230/?arnumber=8294230},
	doi = {10.1109/TPWRS.2018.2807738},
	number = {5},
	urldate = {2025-01-02},
	journal = {IEEE Transactions on Power Systems},
	author = {Höschle, Hanspeter and Le Cadre, Hélène and Smeers, Yves and Papavasiliou, Anthony and Belmans, Ronnie},
	month = sep,
	year = {2018},
	pages = {4819--4830},
}

@article{tirole_incomplete_1999,
	title = {Incomplete {Contracts}: {Where} do {We} {Stand}?},
	volume = {67},
	issn = {1468-0262},
	shorttitle = {Incomplete {Contracts}},
	doi = {10.1111/1468-0262.00052},
	language = {en},
	number = {4},
	urldate = {2025-02-21},
	journal = {Econometrica},
	author = {Tirole, Jean},
	year = {1999},
	pages = {741--781},
}

@article{laffont_using_1986,
	title = {Using {Cost} {Observation} to {Regulate} {Firms}},
	volume = {94},
	issn = {0022-3808},
	url = {https://www.jstor.org/stable/1833051},
	number = {3},
	urldate = {2026-02-17},
	journal = {Journal of Political Economy},
	publisher = {University of Chicago Press},
	author = {Laffont, Jean-Jacques and Tirole, Jean},
	year = {1986},
	pages = {614--641},
}

@misc{australia_energy_market_operator_aemo_2024_2024,
	title = {2024 {Integrated} {System} {Plan} for the {National} {Electricity} {Market}},
	url = {https://www.aemo.com.au/-/media/files/major-publications/isp/2024/2024-integrated-system-plan-isp.pdf?rev=b811f5d66df24e0a980ce0df8eaa5687&sc_lang=en},
	urldate = {2026-02-13},
	author = {{Australia Energy Market Operator (AEMO)}},
	year = {2024},
}

@incollection{pollitt_uncertainty_2025,
	title = {Uncertainty, regulation and the pathways to net zero},
	isbn = {978-1-0353-1435-5},
	url = {https://www.elgaronline.com/edcollchap/book/9781035314355/book-part-9781035314355-22.xml},
	language = {eng},
	urldate = {2026-02-10},
	booktitle = {Handbook on {Electricity} {Regulation}},
	publisher = {Edward Elgar Publishing},
	author = {Pollitt, Michael G. and Duma, Daniel and Covatariu, Andrei},
	month = jun,
	year = {2025},
	note = {Section: Handbook on Electricity Regulation},
	pages = {351--370},
}

@techreport{ofgem_cap_2016,
	title = {Cap and floor regime: unlocking  investment in electricity interconnectors},
	url = {https://www.ofgem.gov.uk/sites/default/files/docs/2016/05/cap_and_floor_brochure.pdf},
	urldate = {2025-11-08},
	author = {{Ofgem}},
	year = {2016},
}

@article{cheng_electricity_2025,
	title = {Electricity {Markets} and {Long}-{Duration} {Energy} {Storage}: {A} {Survey} of {Grid} {Services} and {Revenue} {Streams}},
	volume = {12},
	issn = {2196-3010},
	shorttitle = {Electricity {Markets} and {Long}-{Duration} {Energy} {Storage}},
	url = {https://doi.org/10.1007/s40518-025-00266-9},
	doi = {10.1007/s40518-025-00266-9},
	language = {en},
	number = {1},
	urldate = {2025-07-23},
	journal = {Current Sustainable/Renewable Energy Reports},
	author = {Cheng, Bo and Levin, Todd and Zhou, Zhi and Botterud, Audun},
	month = jun,
	year = {2025},
	pages = {16},
}

@techreport{eso_future_2024,
	title = {Future {Energy} {Scenarios}:  {ESO} {Pathways} to {Net} {Zero}},
	url = {https://www.neso.energy/document/321041/download},
	urldate = {2025-05-27},
	author = {{ESO}},
	month = jul,
	year = {2024},
}

@article{risanger_congestion_2024,
	title = {Congestion {Risk}, {Transmission} {Rights}, and {Investment} {Equilibria} in {Electricity} {Markets}},
	volume = {45},
	issn = {0195-6574},
	url = {https://doi.org/10.5547/01956574.45.1.sris},
	doi = {10.5547/01956574.45.1.sris},
	number = {1},
	urldate = {2025-05-27},
	journal = {The Energy Journal},
	publisher = {SAGE Publications},
	author = {Risanger, Simon and Mays, Jacob},
	month = jan,
	year = {2024},
	pages = {173--200},
}

@techreport{cepa_cap_2025,
	title = {Cap and {Floor} {Regime} for {Long} {Duration}  {Electricity} {Storage}: {Setting} the {Cap} and  {Floor}},
	language = {en},
	author = {{CEPA}},
	month = mar,
	year = {2025},
}

@techreport{ofgem_long_2025,
	title = {Long {Duration} {Electricity} {Storage}: {Technical} {Decision}},
	url = {https://www.ofgem.gov.uk/decision/long-duration-electricity-storage-technical-document},
	language = {en},
	author = {{Ofgem}},
	month = mar,
	year = {2025},
}

@techreport{nsw_department_of_planning_industry_and_environment_long-term_2021,
	type = {Consultation paper},
	title = {Long-{Term} {Energy} {Service} {Agreement} {Design}},
	language = {en},
	author = {{NSW Department of Planning, Industry and Environment}},
	month = aug,
	year = {2021},
}

@article{billimoria_contract_2023,
	title = {Contract design for storage in hybrid electricity markets},
	volume = {7},
	issn = {25424351},
	url = {https://linkinghub.elsevier.com/retrieve/pii/S2542435123002751},
	doi = {10.1016/j.joule.2023.07.002},
	language = {en},
	number = {8},
	urldate = {2025-05-25},
	journal = {Joule},
	author = {Billimoria, Farhad and Simshauser, Paul},
	month = aug,
	year = {2023},
	pages = {1663--1674},
}

@inproceedings{twitchell_targeted_2025,
	title = {Targeted {Financial} {Incentives} for {Long}-{Duration} {Energy} {Storage}},
	url = {https://ieeexplore.ieee.org/document/10891216/?arnumber=10891216},
	doi = {10.1109/EESAT62935.2025.10891216},
	urldate = {2025-02-24},
	booktitle = {2025 {IEEE} {Electrical} {Energy} {Storage} {Applications} and {Technologies} {Conference} ({EESAT})},
	author = {Twitchell, Jeremy and Dave, Prathit and Boff, Daniel and Powell, Devyn and Bhatnagar, Dhruv},
	month = jan,
	year = {2025},
	pages = {1--6},
}

@article{mastropietro_taxonomy_2024,
	title = {A taxonomy to guide the next generation of support mechanisms for electricity storage},
	volume = {8},
	issn = {25424351},
	url = {https://linkinghub.elsevier.com/retrieve/pii/S2542435124001508},
	doi = {10.1016/j.joule.2024.03.015},
	language = {en},
	number = {5},
	urldate = {2025-02-24},
	journal = {Joule},
	author = {Mastropietro, Paolo and Rodilla, Pablo and Batlle, Carlos},
	month = may,
	year = {2024},
	pages = {1196--1204},
}

@article{staadecker_value_2024,
	title = {The value of long-duration energy storage under various grid conditions in a zero-emissions future},
	volume = {15},
	issn = {2041-1723},
	url = {https://www.nature.com/articles/s41467-024-53274-6},
	doi = {10.1038/s41467-024-53274-6},
	language = {en},
	number = {1},
	urldate = {2025-02-05},
	journal = {Nature Communications},
	author = {Staadecker, Martin and Szinai, Julia and Sánchez-Pérez, Pedro A. and Kurtz, Sarah and Hidalgo-Gonzalez, Patricia},
	month = nov,
	year = {2024},
	pages = {9501},
}

@techreport{department_for_energy_security__net_zero_long_2024,
	title = {Long duration electricity storage consultation: designing a policy framework to enable investment},
	language = {en},
	author = {{Department for Energy Security \& Net Zero}},
	year = {2024},
}

@article{dimanchev_consequences_2024,
	title = {Consequences of the missing risk market problem for power system emissions},
	volume = {136},
	issn = {01409883},
	url = {https://linkinghub.elsevier.com/retrieve/pii/S0140988324003475},
	doi = {10.1016/j.eneco.2024.107639},
	language = {en},
	urldate = {2025-02-04},
	journal = {Energy Economics},
	author = {Dimanchev, Emil and Gabriel, Steven A. and Reichenberg, Lina and Korpås, Magnus},
	month = aug,
	year = {2024},
	pages = {107639},
}

@article{rockafellar_optimization_2000,
	title = {Optimization of conditional value-at-risk},
	volume = {2},
	issn = {14651211},
	url = {http://www.risk.net/journal-of-risk/technical-paper/2161159/optimization-conditional-value-risk},
	doi = {10.21314/JOR.2000.038},
	language = {en},
	number = {3},
	urldate = {2025-02-04},
	journal = {The Journal of Risk},
	author = {Rockafellar, R. Tyrrell and Uryasev, Stanislav},
	year = {2000},
	pages = {21--41},
}

@article{shu_beyond_2023,
	title = {Beyond capacity: {Contractual} form in electricity reliability obligations},
	volume = {126},
	issn = {01409883},
	shorttitle = {Beyond capacity},
	url = {https://linkinghub.elsevier.com/retrieve/pii/S0140988323004413},
	doi = {10.1016/j.eneco.2023.106943},
	language = {en},
	urldate = {2024-11-15},
	journal = {Energy Economics},
	author = {Shu, Han and Mays, Jacob},
	month = oct,
	year = {2023},
	pages = {106943},
}

@article{joskow_hierarchies_2022,
	title = {From hierarchies to markets and partially back again in electricity: responding to decarbonization and security of supply goals},
	volume = {18},
	copyright = {http://creativecommons.org/licenses/by-nc-sa/4.0/},
	issn = {1744-1374, 1744-1382},
	shorttitle = {From hierarchies to markets and partially back again in electricity},
	url = {https://www.cambridge.org/core/product/identifier/S1744137421000400/type/journal_article},
	doi = {10.1017/S1744137421000400},
	language = {en},
	number = {2},
	urldate = {2024-06-12},
	journal = {Journal of Institutional Economics},
	author = {Joskow, Paul L.},
	month = apr,
	year = {2022},
	pages = {313--329},
}

@article{roques_adapting_2017,
	title = {Adapting electricity markets to decarbonisation and security of supply objectives: {Toward} a hybrid regime?},
	volume = {105},
	issn = {03014215},
	shorttitle = {Adapting electricity markets to decarbonisation and security of supply objectives},
	url = {https://linkinghub.elsevier.com/retrieve/pii/S0301421517301106},
	doi = {10.1016/j.enpol.2017.02.035},
	language = {en},
	urldate = {2024-06-12},
	journal = {Energy Policy},
	author = {Roques, Fabien and Finon, Dominique},
	month = jun,
	year = {2017},
	pages = {584--596},
}

@article{dowling_role_2020,
	title = {Role of {Long}-{Duration} {Energy} {Storage} in {Variable} {Renewable} {Electricity} {Systems}},
	volume = {4},
	issn = {25424351},
	url = {https://linkinghub.elsevier.com/retrieve/pii/S2542435120303251},
	doi = {10.1016/j.joule.2020.07.007},
	language = {en},
	number = {9},
	urldate = {2024-06-04},
	journal = {Joule},
	author = {Dowling, Jacqueline A. and Rinaldi, Katherine Z. and Ruggles, Tyler H. and Davis, Steven J. and Yuan, Mengyao and Tong, Fan and Lewis, Nathan S. and Caldeira, Ken},
	month = sep,
	year = {2020},
	pages = {1907--1928},
}

@article{sepulveda_design_2021,
	title = {The design space for long-duration energy storage in decarbonized power systems},
	volume = {6},
	issn = {2058-7546},
	url = {https://www.nature.com/articles/s41560-021-00796-8},
	doi = {10.1038/s41560-021-00796-8},
	language = {en},
	number = {5},
	urldate = {2024-06-03},
	journal = {Nature Energy},
	author = {Sepulveda, Nestor A. and Jenkins, Jesse D. and Edington, Aurora and Mallapragada, Dharik S. and Lester, Richard K.},
	month = mar,
	year = {2021},
	pages = {506--516},
}

@book{steven_a_gabriel_complementarity_2013,
	series = {International {Series} in {Operations} {Research} \& {Management} {Science}},
	title = {Complementarity {Modeling} in {Energy} {Markets}},
	volume = {130},
	isbn = {978-1-4419 -6122-8},
	url = {http://pubsonline.informs.org/do/10.1287/0da64ca6-e17a-473d-b876-2df3a5ce2760/full/},
	language = {en},
	urldate = {2024-04-29},
	publisher = {Springer},
	author = {{Steven A. Gabriel} and {Antonio J. Conejo} and {J. David Fuller} and {Benjamin F. Hobbs} and {Carlos Ruiz}},
	year = {2013},
}

@article{ehrenmann_generation_2011,
	title = {Generation {Capacity} {Expansion} in a {Risky} {Environment}: {A} {Stochastic} {Equilibrium} {Analysis}},
	volume = {59},
	issn = {0030-364X},
	shorttitle = {Generation {Capacity} {Expansion} in a {Risky} {Environment}},
	url = {https://www.jstor.org/stable/41316039},
	number = {6},
	urldate = {2024-04-24},
	journal = {Operations Research},
	publisher = {INFORMS},
	author = {Ehrenmann, Andreas and Smeers, Yves},
	year = {2011},
	pages = {1332--1346},
}

@article{ralph_risk_2015,
	title = {Risk {Trading} and {Endogenous} {Probabilities} in {Investment} {Equilibria}},
	volume = {25},
	issn = {1052-6234},
	url = {https://epubs.siam.org/doi/10.1137/110851778},
	doi = {10.1137/110851778},
	number = {4},
	urldate = {2024-04-23},
	journal = {SIAM Journal on Optimization},
	publisher = {Society for Industrial and Applied Mathematics},
	author = {Ralph, Daniel and Smeers, Yves},
	month = jan,
	year = {2015},
	pages = {2589--2611},
}

@article{abada_multiplicity_2017,
	title = {On the multiplicity of solutions in generation capacity investment models with incomplete markets: a risk-averse stochastic equilibrium approach},
	volume = {165},
	issn = {1436-4646},
	shorttitle = {On the multiplicity of solutions in generation capacity investment models with incomplete markets},
	url = {https://doi.org/10.1007/s10107-017-1185-9},
	doi = {10.1007/s10107-017-1185-9},
	language = {en},
	number = {1},
	urldate = {2024-04-23},
	journal = {Mathematical Programming},
	author = {Abada, Ibrahim and de Maere d’Aertrycke, Gauthier and Smeers, Yves},
	month = sep,
	year = {2017},
	pages = {5--69},
}

@article{mays_asymmetric_2019,
	title = {Asymmetric risk and fuel neutrality in electricity capacity markets},
	volume = {4},
	copyright = {2019 The Author(s), under exclusive licence to Springer Nature Limited},
	issn = {2058-7546},
	url = {https://www.nature.com/articles/s41560-019-0476-1},
	doi = {10.1038/s41560-019-0476-1},
	language = {en},
	number = {11},
	urldate = {2024-04-22},
	journal = {Nature Energy},
	publisher = {Nature Publishing Group},
	author = {Mays, Jacob and Morton, David P. and O’Neill, Richard P.},
	month = nov,
	year = {2019},
	pages = {948--956},
}

@article{mays_financial_2023,
	title = {Financial {Risk} and {Resource} {Adequacy} in {Markets} {With} {High} {Renewable} {Penetration}},
	volume = {1},
	copyright = {https://ieeexplore.ieee.org/Xplorehelp/downloads/license-information/IEEE.html},
	issn = {2771-9626},
	url = {https://ieeexplore.ieee.org/document/10273857/},
	doi = {10.1109/TEMPR.2023.3322531},
	language = {en},
	number = {4},
	urldate = {2024-04-22},
	journal = {IEEE Transactions on Energy Markets, Policy and Regulation},
	author = {Mays, Jacob and Jenkins, Jesse D.},
	month = dec,
	year = {2023},
	pages = {523--535},
}

@IEEEtranBSTCTL{IEEEexample:BSTcontrol,
  CTLuse_url = "no",
  CTLuse_doi = "no",
  CTLuse_eprint = "no"
}

\end{document}